# Mind the Gap: A Framework (BehaveFIT) Guiding The Use of Immersive Technologies in Behavior Change Processes

**Wienrich, Carolin[1*], Döllinger, Nina, Ines[1], Hein, Rebecca[1]**

[1]Human-Technique-Systems, Institute of Human-Computer-Media, faculty of human sciences, University of Würzburg, Würzburg, Germany

**\* Correspondence:**
Corresponding Author
carolin.wienrich@uni-wuerzburg.de

## Abstract

The design and evaluation of assisting technologies to support behavior change processes have become an essential topic within the field of human-computer interaction research in general and the field of immersive intervention technologies in particular. The mechanisms and success of behavior change techniques and interventions are broadly investigated in the field of psychology. However, it is not always easy to adapt these psychological findings to the context of immersive technologies. The lack of theoretical foundation also leads to a lack of explanation as to why and how immersive interventions support behavior change processes. The Behavioral Framework for immersive Technologies (BehaveFIT) addresses this lack by (1) presenting an intelligible categorization and condensation of psychological barriers and immersive features, by (2) suggesting a mapping that shows why and how immersive technologies can help to overcome barriers, and finally by (3) proposing a generic prediction path that enables a structured, theory-based approach to the development and evaluation of immersive interventions. These three steps explain how BehaveFIT can be used, and include guiding questions and one example for each step. Thus, the present paper contributes to guidance for immersive intervention design and evaluation, showing that immersive interventions support behavior change processes and explain and predict 'why' and 'how' immersive interventions can bridge the intention-behavior-gap.

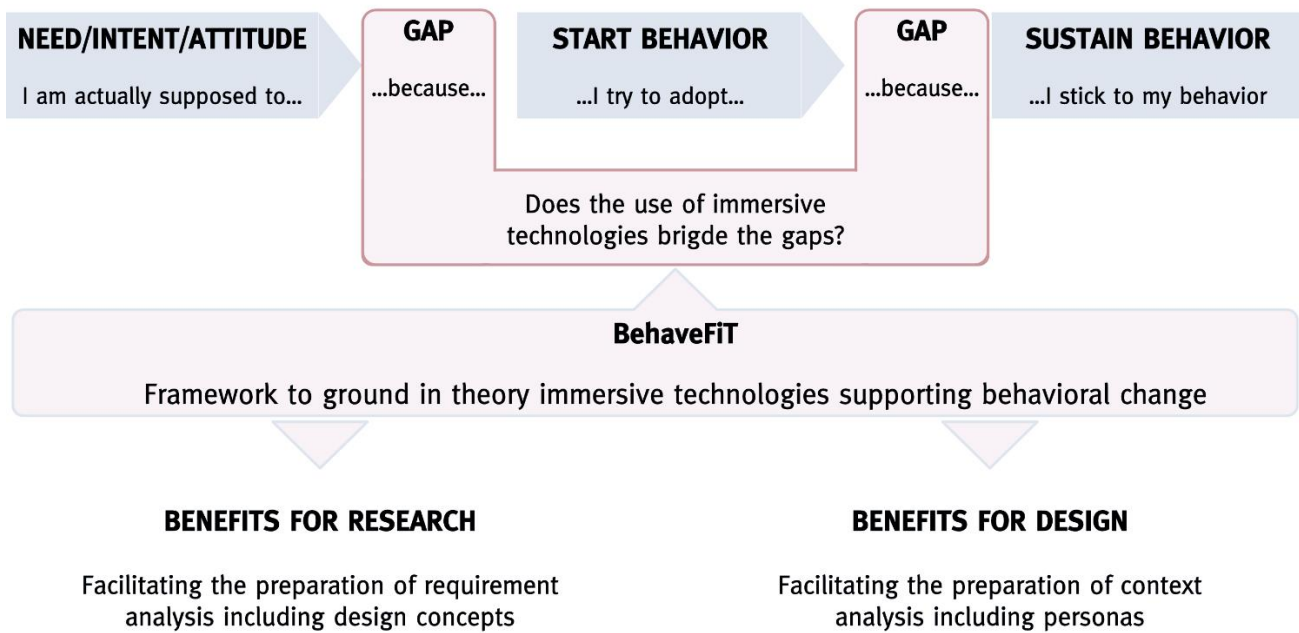

**Figure 1: The teaser figures shows the contribution of the Behavioral Framework of Immersive Technologies - BehaveFIT**

## 1 Introduction

"The road to hell is paved with good intentions." We all know that the translation from intending to change a behavior to successfully changing it often fails. We all have experienced what researchers call the intention-behavior-gap, value-behavior-gap, attitude-behavior-gap, or the knowledge-behavior-gap (Ajzen, 2011; Carrington et al., 2010; Gollwitzer & Bargh, 1996; Sheeran, 2002; Whitmarsh, 2009) (first row of Figure 1).

While intentions may be the best predictor of behavior, on average they account only for 28% of the variance in future behavior, connoting that there may be other factors that predict or rather inhibit a successful behavior change (Sheeran, 2002).

Indeed, the field of psychology has a long tradition of explaining 'why' the gap occurs, including a wide range of behavior determinants (e.g., Davis et al., 2014; Gifford et al., 2018; Kollmuss & Agyeman, 2002). However, the extensive quantity of determinants operating on different levels may impede the design, the evaluation, or the comparison of intervention applications, particularly for researchers from other fields. Since behavior change is challenging, assisting technologies that strive to support behavior change, have become an essential line of research within the human-computer interaction (HCI) research in general (Klasnja et al., 2011; Stawarz et al., 2015), and within the field of immersive interventions (mainly virtual reality applications) in particular (e.g., Kuo, 2015; Riva et al., 2019; Sherman & Craig, 2018; Sun Joo Grace Ahn, 2016). After referring to the user's perceptions, the immersive technologies' features reveal many possibilities that could support behavior change. However, the description of why these features and possibilities may facilitate behavior change processes and how they are related to the determinants known from psychology often remains phenomenological (Riva et al., 2019; Stawarz & Cox, 2015). Accordingly, the lack of transfer between the disciplines results in a lack of predicting models between the features and possibilities of immersive technologies and the psychological behavior determinants, which further results in a lack of theoretical based prediction for the design and evaluation of immersive interventions.

In the present paper, we aim to use research from other disciplines to explain why and how immersive interventions support behavior change processes. According to the claim for theoretical research contributions raised by Wobbrock & Kientz (2016), we present a framework that targets (1) the summation of a variety of models describing and predicting behavior change from a psychological perspective. The summation results in an intelligible categorization of determinants and barriers. It further targets (2) the summation of the most common features and possibilities of immersive technologies from the human-computer interaction (HCI) perspective. This summation focuses on features of immersive technologies that might be relevant to support behavior change processes. While the first two aims contribute to the why, our final scope is (3) to form generic predicting paths that show how the immersive features and possibilities affect the psychological determinants and how they impact behavior change processes. Thus, both summation results will be mapped on to each other - the features and possibilities of immersive technologies will be mapped on to the categorization of the psychological determinants and barriers.

Consequently, the framework enables theory-driven predictions for the designing and evaluating of immersive intervention technologies designed to support behavioral changes. Figure 1 illustrates the value of the framework    when immersive technology has been applied to bridge the intention-behavior-gap.

To achieve these aims, we first provide an overview of the psychological research that explains why the intention-behavior-gap occurs by introducing barriers that impede the adoption of new behavior. In the next step, we introduce intervention mechanisms based on so-called behavior change models that aim to explain how the intention-behavior-gap can be overcome. The psychological introduction is followed by the summation of research from the field of HCI, focusing on immersive interventions. Hence, thirdly we discuss the potentials of different features of immersive technologies that indicate 'why' they can support behavior change. Finally, we illustrate the currently available empirical indicators of 'how' immersive technologies can support behavior change. After the theoretical overview (i.e., summations), we introduce the framework BehaveFIT, including the developing methods and results referring to the theoretical section. The benefits and limitations of the framework are discussed by considering the design and evaluation of immersive interventions supporting behavioral changes.

## 2    Related Work

## 2.1    THE PSYCHOLOGY OF WHY THE INTENTION-BEHAVIOR-GAP OCCURS

While addressing slightly different barriers, the intention-, value-, attitude- or the knowledge-behavior-gap are roughly summarized as the intention-behavior-gap as follows. The intention-behavior-gap describes the failure of the translation from intending to change a behavior to successfully changing it (Ajzen, 2011; Carrington et al., 2010; Gollwitzer & Bargh, 1996; Kollmuss & Agyeman, 2002; Schwarzer, 2014; Sheeran, 2002; Whitmarsh, 2009). The focus of psychological research on the intention-behavior-gap depends on the respective behavioral domain. Nonetheless, some generally valid barriers explaining this failure have been identified. The current section gives an overview of them but raises no claim of completeness. For a comprehensive presentation of the variety of the psychological barriers, we further refer to the work of Gifford et al. (2018), Cane et al. (2012), Blake (1999) and Kollmuss & Agyeman (2002).

Blake (1999) has identified three barriers; individuality, responsibility, and practicality. Individuality refers to barriers lying within the person such as attitudes, personality traits, or predispositions. Responsibility, or rather irresponsibility, refers to the lack of control or perceived influence on the situation. It correlates to distrust or disbelief in the need for change or a low expectancy of self-efficacy. Practicality refers to constraints lying outside the person, including limited resources or facilities, for example.

Other researchers have identified demographic factors, external factors, and internal factors that prevent people from changing their behavior (Kollmuss & Agyeman 2002). Demographic factors include gender and education. External factors entail institutional (e.g., infrastructure to use public transport to change one's environmental behavior), economic (e.g., monetary benefit), social, and cultural factors (e.g., social norms). Internal factors include motivation (primary motives like altruism or immediate selective motives like comfort), value (that are shaped by, e.g., experiences, role models, education), knowledge, attitude (positive or negative valence toward the behavior), awareness (that underlies cognitive limitations, for example negative consequences are often slow and gradual, and thus not immediately tangible), emotional involvement (e.g., reactance, reduction of dissonance, feeling of helplessness), locus of control (perceived ability to influence the behavior), and responsibility and priorities (e.g., conflicts between contradicting wishes).

More recently, (Gifford et al., 2018) have identified six categories of psychological barriers, including most of the categories mentioned above: no need to change, conflicting goals and aspirations, interpersonal relations, government and industry, tokenism, and lacking knowledge. No need to change describes the conviction that a behavior change is unnecessary due to denying the problem or believing it will resolve itself. Gifford et al. (2018) subsumed several barriers in this category: disbeliefs    include mistrust, denial, and reactance. Broader belief systems (ideologies) include technosalvation, suprahuman control, and system justification. Limited cognition includes a lack of self-efficacy, optimism bias, and confirmation bias. Conflicting goals and aspirations represent conflicts between goals related to the intended behavior change and other goals, and encompasses costs, including financial investment or behavioral momentum, and perceived risks like unprofitable time investment. Interpersonal relations include the fear that significant others will disapprove of the changed behavior and subsume barriers related to social norms, social comparison, and social risks. Government and industry represent the lack of responsibility of individuals due to the belief that the industry or the government should take charge and include, among others, the perceived program inadequacy barrier. Tokenism refers to the beliefs of people who already adopted some steps of the intended behavior change and believe that they have done enough. While tokenism itself describes that easy changes tend to be chosen over action with higher effort, the rebound effect involves diminished or erased actions by subsequent actions. Lacking knowledge describes when people are paralyzed by their lack of knowledge about selecting the correct action, how to undertake the action, and the relative benefits of different actions. Gifford et al. (2018) subsumed barriers stemming from limited cognition like ignorance or numbness; and perceived risks like functional risks.

**Desideratum 1 for the framework development.** The psychological barriers explain 'why' the *intention-behavior-gap* occurs. Perceived distances to approval or role models, and especially to tangible consequences, prevent people from starting or sustaining behavior change processes. However, although we already presented a condensed overview, it is hard to find starting points to design or evaluate immersive interventions from existing classifications and descriptions. Consequently, we identify the need for an intelligible categorization of those barriers to provide a guidance, particularly for non-psychologists.

## 2.2 THE PSYCHOLOGY OF HOW THE INTENTION-BEHAVIOR-GAP CAN BE OVERCOME

The mere description of why does not allow for predicting how the interventions deploy their effectiveness. Hence, this section sums up the psychological models and taxonomies that include the relations between the barriers, their occurrences within the change process, or milestones that accompany a successful change process. Theory-driven approaches based on so-called behavior change models and structured taxonomies contribute to improved research on behavioral interventions (see also guidelines of Byrne, 2020). Because of this, those approaches seem to be a promising starting point for improvement of immersive interventions. Note again, our selection raises no claim to completeness.

### 2.2.1 Behavior Change Models

Davis et al. (2014) provided a review of different theories of behavior change. Among the most frequently cited theories are The Health Belief Model (HBM, Becker & Maiman, 1975), the Theory of Planned Behavior (TPB, Ajzen, 1980), the Transtheoretical Model (TTM, Prochaska & Velicer, 1997), the Social Cognitive Theory (SCT, Bandura, 1977), and the Health Action Progress Approach (HAPA, Schwarzer, 2014).

**Fear Appeal Models.** Fear appeal determines behavior change mainly by the perceived threat of consequences of the original behavior linked directly to a behavior change. One of the essential models in fear appeal models is the HBM (Becker & Maiman, 1975). Here, the fear of consequences is linked with the weighing up of costs and benefits of a potential behavior change and some demographic variables.

**Motivational Models.** The area of motivational models goes beyond the idea of fear as a single motivator for behavior change. They include a variety of the barriers mentioned in section above and link them to behavior change. For example, the TPB (Ajzen, 1980) introduced the perceived behavior control, i.e., the perceived difficulty to perform the new behavior, the perceived behavior norms, and personal behavior attitudes as motivators and barriers in behavior change. Similarly, the SCT (Bandura, 1977) introduced motivators that contribute to behavior change initiation. The SCT emphasizes the social context and their self-efficacy expectations for implementing the behavior change. Motivational models focus on the intention to change behavior and the initiation of new behavior. They mainly illustrate a connection of motivators to behavior change rather than offering mechanisms for long-term implementation of change or the development of coping strategies with relapses.

**Stage models.** Addressing the problem of initiation and maintaining new behavior, stage models divide the behavior change process into a pre-defined number of behavioral changes. The stages contain specific mechanisms contributing to the transition between stage and classifications of persons regarding their progress. For example, the TTM (Prochaska & Velicer, 1997) describes six stages: precontemplation (awareness-raising), contemplation (decision-oriented), preparation (preparing the initiation of the target behavior), action (initiation of the target behavior), maintenance (maintaining the target behavior over time) and termination (automatic execution of the behavior). A drawback of these stage models is that the strict division into stages suggests a linear temporal course that does not reflect real behavior changes. Additionally, the length of the stages (e.g., maintenance = 6 months) seems arbitrary and could not be verified empirically (Bridle et al., 2005).

**Integrative Models.** Finally, some models go beyond pure motivation, which overcomes the fixation on development stages. Besides the expectation of results and risk perception, the HAPA (Schwarzer, 2014), for example, emphasized the role of self-efficacy expectation (similar to the SCT). However, the model goes beyond the formation of intentions. It also includes different types of self-efficacy during action and coping, planning and behavior implementation, and evaluation. Besides, the barriers mentioned above and behavior change are included as risk factors in the transition from planning to action maintenance. Like the stage models, a definition of user groups is possible (Non-Intender, Intender & Active). However, the change between the stages is less arbitrarily defined and instead depends on various personal and situational variables forming the motivation within each stage and allowing for relapses.

### 2.2.2 Taxonomy of Behavior Change Techniques

Additional support in the development of behavioral evaluations is provided by the "Taxonomy of Behavior Change Techniques" by Michie et al. (2013). This taxonomy comprises a summary of 93 types of behavior change techniques, which are grouped into 16 categories. The categories of behavior change techniques proposed by Michie et al. (2013) are scheduled consequences, reward and threat, repetition and substitution, antecedents, associations, covert learning, natural consequences, feedback and monitoring, goals and planning, social support, comparison of behavior, self-belief, comparison of outcomes, identity, shaping knowledge, and regulation.

In addition to the psychological barriers describing the why, this classification gives an overview of how existing interventions work. Hence, it offers the possibility for immersive interventions to refer to already empirically tested behavioral techniques. For example, to address a psychological barrier in the context of interpersonal relations, a comparison between interventions with a focus on social support or comparison of behavior might be appropriate to refer to. To overcome the barrier of the lack of self-efficacy expectation, surveying interventions in the category of self-belief or identity could be supportive.

### 2.2.3 Identification of Interactions and Moderators

To correctly interpret the outcome of an intervention, it is necessary to consider the role of confounding variables moderating intervention effects. Byrne (2020) roughly groups such moderators into three categories: temporal intervention context, mode of delivery, and individual characteristics. Inspired by a commentary by Scholz (2019), the temporal context of an intervention has been addressed by several researchers (e.g., Bolger & Zee, 2019; Rothman, 2019) in recent years. Discussions include, amongst others, the time of day, the time of year or time within the process of behavior change that the intervention is applied, or the cycle of repetitions. Additionally, the mode of delivery (Wilson, et al., 2020) or the spatial context can also influence the outcome of a behavior change intervention. These variables include the choice of medium compared to in-person interventions that might be particularly relevant for immersive interventions. The third group of moderators is the characteristics of the target person (Byrne, 2020). These include both individual personality traits and the socio-ecological characteristics of a person. They can be controlled by carefully defining the intervention's target group (see also Wienrich & Gramlich, 2020).

> **Desideratum 2 for the framework development.** Several models predict behavior change processes, thus guiding the design of interventions, identifying influencing factors, controlling moderators and confounding variables, and defining target groups. Taxonomies of behavior change techniques offer additional classification and orientation. However, immersive interventions are rarely based on those models, and taxonomies doubting their quality (Byrne, 2020). Consequently, we identified the need for a predictive framework linking immersive interventions to the theoretical base of behavior change.

## 2.3 THE POTENTIALS OF IMMERSIVE TECHNOLOGIES - WHY CAN THEY SUPPORT BEHAVIOR CHANGE?

Immersive technologies have many potential ways to support behavior change processes since they refer to perceptions of users. The following section summarizes the most promising features considering behavior change and the corresponding perceptional links (see Table 1).

### 2.3.1 Immersion and Presence

Beyond doubt, the terms immersion and presence are inextricably linked with immersive technologies. In contrast to other fields of research (e.g., media communication), a long debate in the field of HCI ended in the agreement that immersion stands for what the technology delivers in all sensory and tracking modalities and that it can be objectively assessed. In contrast, presence can be defined as a human reaction to a system of a certain level of immersion and thus describes a subjective state (Slater, 1999, 2003). Further research identified different aspects of presence such as social presence (explained below), telepresence, spatial presence or being there, physical presence, self-presence, or place illusion, and plausibility illusion (for a systematic review, see Skarbez, Brooks, & Whitton, 2017). Immersion and (spatial) presence could probably be considered as so-called "hygiene factors" that are necessary to a certain extent to allow other potentials of immersive technologies (e.g., self-presentation) to become effective, but whose occurrence alone does not lead to the intended behavioral effects. It might be conceptualized similarly to the role of pragmatic quality within the field of user experience (Hassenzahl et al., 2010).

### 2.3.2. Self-Representation and the illusion of virtual body ownership.

One exclusive feature of VR is the sense of embodiment elicited by the appearance of a virtual alter ego – an avatar. Much research has shown that the virtual embodied representation can influence the user's self-perception (see 2.3.1). The corresponding perceptions are mainly based on the illusion that the virtual representation is part of one's physical body, called the sense of embodiment (Kilteni et al., 2012) or the illusion of virtual body ownership (IVBO). The sense of embodiment occurs with the corresponding phenomenological senses of self-location: the spatial experience of being inside a body, agency: the experience of controlling a body, and ownership: self-attribution of a body (Kilteni et al., 2012). Recently, the sense of change was added to describe further the IVBO (Roth, 2019). The IVBO depends on anthropomorphism and the predictability and synchronicity of the avatar's movements regarding the user's physical movements (Maister, Slater, Sanchez-Vives, & Tsakiris, 2015). One behavior-relevant phenomenon that is closely related to the effects of virtual self-representation is the Proteus effect. It describes a change in behavior according to behavior users attributed to their virtual representation, even after they left the virtual simulation (Ratan et al., 2020; Yee & Bailenson, 2007).

The potential of self-representation and particularly the corresponding sense of embodiment or rather the illusion of virtual body ownership refers to personalizing of any (immersive) experience or making any (immersive) experience literally tangible.

### 2.3.2 Other- and Context-Representation

Similarly to self-representation, other users can be virtually embodied and varied in their appearance and behavior as social partners. The representation of real or artificial others gains in relevance when users start to interact with virtual others. Consequently, not only the appearance and behavior of interaction partners are crucial for a possible behavior change, but also the representation of interaction or social context (e.g., number of interaction partners, presenting similarities or differences, features of group processes, social status, social interdependence) (Wienrich et al., 2018). Similarly to the definition of spatial presence as a consequence of immersion, others' inclusion and representation can lead up to a sense of social presence or co-presence. Social presence can be defined as the "sense of being with another" (Biocca & Harms, 2002, p. 456). In the context of assessing the social presence, the sub-dimensions of empathy, involvement, and feelings have been described (De Kort et al., 2007). The sense of social presence mainly depends on sensory factors (e.g., visual representation of others, interactivity), or on the social context (e.g., physical proximity, social cues) (Oh et al., 2018) as well as on the task-related social interdependence between the actors (e.g., avatars) (Wienrich et al., 2018). The representation of others refers to the potential of regulating and simulating social aspects within a behavior change process.

### 2.3.3 Virtual Objects and Environmental Cues

The design space of the situational context in VR is almost limitless. Any object, any feature of objects and any relation between objects and between objects and persons are modifiable. In particular, decoupling time and space from natural physical laws has been identified as a supportive aspect of behavioral change (Ahn et al., 2015; Sherman & Craig, 2018). For example, consequences can be directly elucidated, and users can be transported to any location or into (or out of) any context.

Potentials referring to manipulating the object-representation and the environment's representation are promising features to influence spatial or temporal aspects and factors of context on behavior change.

**Table 1**: **Features of immersive technologies and corresponding perceptions**

| Promising Features | Corresponding Perceptions |
| --- | --- |
| immersion | sense of presence (e.g., spatial presence, self-presence) |
| modifiable self-representation | self-perception (e.g., IVBO, sense of agency, sense of change) |
| modifiable other- and context-representation | other- and context-perception (e.g., sense of social presence) |
| modifiable virtual objects and virtual environments | various perceptions (e.g., sense of time, sense of space) |

> **Desideratum 3 for the framework development.** Immersive technologies (mainly investigated in VR) show features with a high potential to support the behavioral change process. The corresponding perceptions occurring during or after an immersive experience seem promising to design personalized and socially relevant interventions overcoming psychological barriers preventing people from starting or sustaining behavior change processes. However, these features are rarely related to the psychological barriers leading up to a lack of clarity about whether, why, and how they affect human behavior change. Consequently, we identified the need for relating the features and the corresponding perceptions of immersive technologies to the psychological barriers impeding behavior change.

## 2.4 EMPIRICAL INDICATORS OF HOW IMMERSIVE TECHNOLOGIES CAN SUPPORT BEHAVIOR CHANGE

This section sums up empirical results indicating how immersive interventions support behavior change processes. Since no models or taxonomies exist in this area, our report relies on empirical results.

A large proportion of investigations on immersive technologies and behavior change address the use of virtual exposure therapy in the treatment of clinical phobias or use exposure therapy to treat addictions (for reviews see: Morina, et al. (2015), Powers & Emmelkamp (2008)). The main focus of these applications is on coping with negative emotions or addictive cravings that are triggered by specific stimulus situations. The further potential for immersive technologies to promote new behavioral patterns is not addressed in these therapeutic applications.

However, recent studies have examined how the potential of immersive technologies mentioned above can be used to induce behavior change in non-clinical settings.

### 2.4.1 The effect of immersion on behavior change.

Some studies focus on the feature of immersion itself as a trigger for behavior change. For example, Ahn, Bailenson, & Park (2014) examined the extent as to which the immersion of an intervention technology affects pro-environmental behavior intentions. The results revealed higher environmental behavior intentions in a highly immersive VR condition than a lower immersive printed or 2D video condition. A follow-up study demonstrated that the environmental intentions correlated with the degree of interactivity within an immersive application (Ahn et al., 2015). To further identify how immersion impacts behavior change, some scientists consider the mediating role of corresponding perceptions mentioned above. One example of such an approach is the third series of experiments by Ahn et al. (2016). The studies embodied people in different virtual animals and revealed that presence did not have a mediating effect between immersion and behavior change, but the illusion of virtual body ownership. Similar results have been shown in other studies (Ahn et al., 2019; Herrera et al., 2018).

We can learn from these and similar studies that immersion is one crucial feature. However, other features and corresponding perceptions also explain important variance of the effects of immersive interventions on behavior change.

### 2.4.2 The effect of self- or context-presentation on behavior change.

Other studies addressed the feature of self-presentation or context-representation on behavior change. Fox & Bailenson (2009), Kim et al., (2014), and Kuo (2015) provided feedback on behavioral consequences presented on one's virtual self-representation. Fox & Bailenson (2009) showed that an embodied avatar could increase behavioral intentions in exercising behavior when it is modified according to the user's current exercising performance. Also, Kim et al. (2014) and Kuo (2015) showed that modified virtual self-representation could significantly influence the motivation to change one's health behavior.

Additional to feedback on their avatar, the virtual environment can impact behavioral change. In an experiment by Soliman, Peetz, & Davydenko (2017), the characteristics of the presented environment, either as natural or artificial, impacted pro-environmental attitudes. In the natural environment, the users felt more connected to nature and pro-environmental behavior than in a virtual urban environment. Likewise, Ahn et al. (2015) showed that behavior is influenced by the environmental framing of an immersive message. Mediated by increased self-efficacy, using positive or negative environmental cues significantly impacted behavioral intentions, with positive framing leading to the higher intention of paper saving behavior than negative. Other experiments, like the work of Hsu et al. (2018) showed that feedback on the impact of behavior patterns on the environment can help to motivate behavioral intentions.

The research on one's self- or context-representation shows that besides the degree of immersion, other modifiable features such as the type of feedback or the framing of the situation are significant modifiers to support behavior change.

### 2.4.3 Transfer between virtual and real behavior

A frequently addressed problem in research on the effectiveness of clinical VR exposure therapy is whether the results of virtual intervention can be transferred to perceptions and behavior in the non-virtual world (i.e., the real world). The field of exposure therapy revealed several results that show a transfer of these effects, both in the short-term and the long-term after leaving the immersive environment (Morina et al., 2015; Powers & Emmelkamp, 2008). In addition to exposure therapy, the Proteus effect (described above) indicates that behavioral experiences during the immersive experience are also transferred to real behavior in both the short term and the long term after the experience (see for a recent review: Ratan, et al., 2020).

However, in line with Stawarz & Cox (2015), some results also show the limitation of being more motivated than truly exhibiting changed behavior. For example, Kim et al. (2014) found that the positive effects of immersive interventions rapidly declined when the immersive system was not available for further exercises. The authors drew the conclusion that immersive interventions had an initial motivating effect but not a sustaining effect on real behavior.

---

**Desideratum 4 for the framework development.** Many studies from different behavioral domains (e.g., therapy, environment, health) show that immersive interventions can support behavior change. The empirical results revealed that different features manipulating immersion can influence the behavior change, that besides immersion, other features take effect on behavior change, and immersive interventions can be transferred to real behavior after the intervention. However, some studies presented conflicting results (despite the fact, we are not considering the publication bias). Further, only few of them provide detailed information of the why and how by linking the features with corresponding perceptions and psychological determinants and barriers. Consequently, we identified a need for predicting more precisely why and how immersive interventions help to overcome psychological barriers and support behavior change.

---

## 2.5   OUTLINE OF PRESENT WORK

The related work revealed plenty of research on psychological barriers, behavioral models, and the potentials of immersive interventions to support behavior change processes. Four desiderata corroborate the need for an intelligible summation and linkage between the psychology of why and how and the why and how of immersive interventions. To address the desiderata, we present a re-categorization of common barriers of behavior change from psychological research on *why* the intention-behavior-gap occurs. The result is an intelligible categorization of determinants and barriers (refer to desideratum 1). The mapping between behavior-relevant features of immersive technologies on to these categories provides an overview of possible immersive interventions (refer to desideratum 3). We then address the summation of psychological models on *how* behavior change techniques operate and research *how* immersive technologies affect behavior change (refer to desideratum 2). Finally, our framework provides a generic prediction path that enables a structured, theory-based approach to develop and evaluate immersive interventions (refer to desideratum 4).

## 3    Method

The framework was developed in an iterative process, including literature reviews and expert feedback. A short outline of this process follows.

### 3.1    Literature Review

We reviewed literature stemming from the field of psychology and HCI. The procedure followed the guidelines of Baumeister & Leary (1997) and Moher et al. (2015). The result was (1) an overview of behavior change barriers (refer to the psychology of why) and (2) psychological behavior change models and taxonomies (refer to the psychology of how). Furthermore, it resulted in (3) an overview of immersive features (refer to the potentials, the why, of immersive technologies) and (4) their effects on behavior change (refer to the empirical indicators, the how, of immersive technologies). The analysis started with the summation of results stemming from the literature by applying the Affinity Diagram method (Holtzblatt & Beyer, 2017) in an internal workshop. By using the Affinity Diagram method, data can be classified based on their relationships. Our analysis followed the usual Affinity Diagram process, which includes five steps: (1) Collecting abstract data onto post-it notes; (2) Clustering post-its by answering the question: Is the post-it similar or different to the existing groups of post-its? (3) Naming the clusters; (4) Ranking the clusters; (5) Finding relationships between clusters and naming them. After the summation of literature review data, we incorporated additional aspects stemming from the discussion in the affinity workshop. The affinity workshop led to the first draft of the framework that has been evaluated and improved by one focus group and two expert interviews.

### 3.2    Focus Group and Expert Interview

A focus group comprising researchers and practitioners working in different fields, including psychology, media sciences, and HCI (Kuniavsky, 2003) evaluated the first draft. The main aim was the evaluation of the intelligibility of the framework. Without prior explanation, participants were given a visualization of the framework. After 15 minutes of discussion in sub-groups of four, we introduced the main ideas of the framework. The final group discussion revealed parts that were coherent and parts that were barely intelligible. To capture diverse expertise on the theoretical foundation, completeness, and value of the framework, we conducted two expert interviews. We interviewed a professor of social and behavioral psychology who had been working as a psychological researcher since 1990 and one professor of HCI research with a focus on VR and immersive technologies who had been working in this field since 1996. Both rated their expertise in the respective field as high (4) or very high (5) on a 5-pt Likert scale. At the beginning of each interview, we introduced the main ideas of the framework. In the following 45-minute interview, the psychological expert was mainly interviewed about the incorporation of the psychological models and determinants within the framework. The interview with the VR/HCI expert focused on the inclusion of the features and mechanisms of immersive technologies.

The focus group's statements and the answers from the two experts were categorized via a structured content analysis into different thematic areas: labeling, framework structure, expansion, and tangibility. Based on the focus group and expert interviews, we revised the first draft leading up to the version presented in the following section.

## 4    BehaveFIT

### 4.1    BehaveFIT: Why Can Immersive Technologies support Behavior Change

This section addresses desiderata 1 and 3 by categorizing the psychological barriers and by mapping the features and corresponding perceptions of immersive technologies on to this categorization. The categorization and mapping contribute to the question: Why can immersive Technologies support behavior change?

#### 4.1.1 BehaveFIT: Intelligible Categorisation of Psychological Barriers

Section 2.1 introduces psychological barriers explaining why the intention-behavior-gap occurs. To approach the desideratum 1 that identified the need for an intelligible categorization of those barriers to provide guidance for intervention design and evaluation, BehaveFIT suggests a categorization that includes six distances impeding old behavior detachment or the adoption of new behavior. In the following, the distances are introduced. Figure 2 (first column) shows which psychological barriers (2.1) from the literature belong to the categories.

**Temporal distance** describes barriers referring to the tendency to place a higher subjective value on rewards received immediately than those received in the future. Hence, the temporal distance of risks or benefits that often occur far in the future can lead to a failure to change behavior.

**Spatial distance** describes barriers referring to consequences that are geographically far away from the target person. Far away consequences are perceived as subjective, less affecting, consequential, or present, and thus, they are easier to ignore.

**Distance from benefits** describes barriers referring to the required effort in relation to the expected benefit of the change. It is related to the lack of immediate or near consequences but focuses on individual costs and risks associated with the change behavior.

**Distance from effect/from control** describes barriers referring to the lack of perceived control and self-efficacy. This distance can occur due to individual constraints or the expectation that other (more powerful) forces should solve the problem.

**Distance from information** describes barriers referring to a lack of information or to biased information processing. The lack of information interferes with problem discernment, and biased information processing correlates with justification behavior - both important for a successful change process.

**Distance from persistence** describes barriers referring to the way people are reluctant to change their behavior, particularly their habits. Behaviors, easy to adopt, tend to be chosen over action with higher effort cost, although the impact is relatively small.

### 4.1.2 BehaveFIT: Mapping of Features and Corresponding Perceptions to the Psychological Barrier Categories

Section 2.3 introduces features and corresponding perceptions of immersive technologies, indicating why they can support behavior change processes. To approach the desideratum 3 that identified the need for relating the features and the corresponding perceptions to the psychological barriers impeding behavior change, BehaveFIT suggests a mapping on to the six barrier categories described above. The mapping shows where immersive technologies can help to overcome the barriers. In the following, the mapping is introduced. Figure 2 (second, third, and fourth column) shows which features from the literature belong to the mapping.

Several features of immersive technologies manipulate the sense of time and corresponding effects and thus approach barriers referring to the category of temporal distance. Far away consequences (benefits and risks) can be brought nearer to the current behavior by showing the future self or future environments. Vice versa, users can experience events from the past to gain a better understanding. The corresponding modified self-or context perceptions can facilitate behavior change processes.

One crucial feature of immersive technologies is the ability to modify the sense of space, addressing barriers referring to the category of spatial distance. Again, far away consequences can be brought nearer to the current behavior. Particularly, other places and other persons can be experienced tangibly. The teleportation of oneself to other places or into other bodies or vice versa - bringing other places or persons into their own backyard – can facilitate behavior change processes.

Several features of immersive technologies increase the perception of benefits and thus approach barriers referring to the category of distance to benefits. Similarly to change processes themselves, rewarding effects often occur far in the future while costs occur immediately, fostering the focus on an over-proportional perception of effort. Modified immersive self or environments and corresponding perceptions can change the focus to the benefits and facilitate behavior change processes.

Particularly, modifying self-representation and the corresponding self-perception addresses barriers referring to the category of distance from effect/control by increasing the feeling of control and self-efficacy. The Proteus effect has proven that the effects go beyond mere exposure situations and manifest in real behavior habits, thus facilitating behavior change processes.

Augmenting or explaining complex information can facilitate information processing or the unmasking of biased information processing. Modifiable virtual objects and environments can support the information presented in addition to the behavior change processes.

Motivating repetitions by modified environments and gamification elements (e.g., a beach instead of the hospital), and increasing the feeling of relatedness by virtual companions are examples of approaching the distance to persistence by immersive technologies. Particularly, mapping on to the lack of motivation can facilitate the maintenance of behavior change processes.

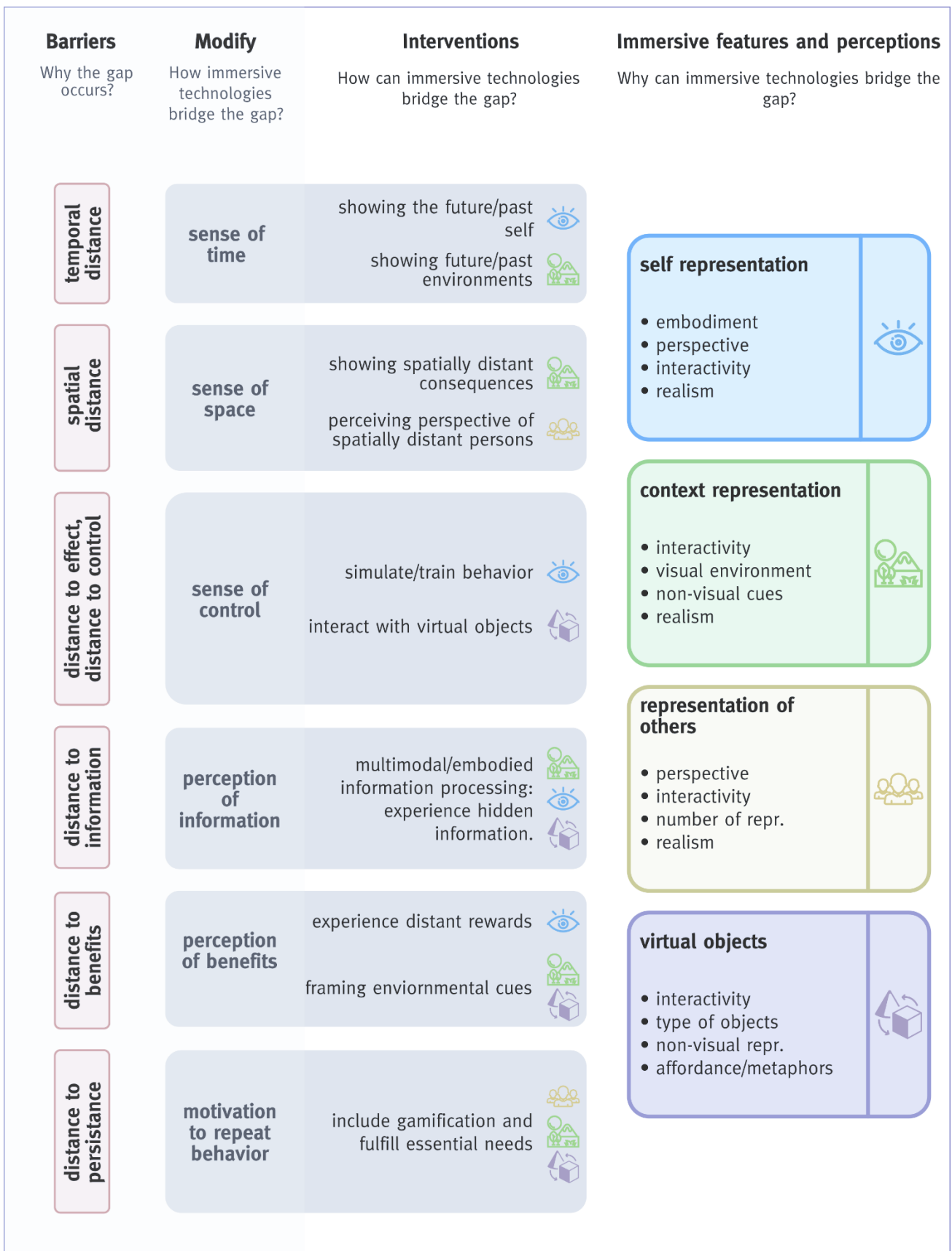

**Figure 2: Sums up the framework BehaveFIT explaining why and how immersive technologies support behavior change processes.**

## 4.2 BehaveFIT: How Do Immersive Technologies Support Behavior Change

Section 2.2 introduces psychological models and taxonomies explaining how interventions should be designed to overcome the psychological barriers and support behavior change. This section addresses desideratum 2 and 4 that identified the need for a predictive framework linking immersive interventions to the theoretical base of behavior change. The linkage can guide the design and evaluation of immersive interventions. Consequently, this section introduces three steps explaining how BehaveFIT can be used, including guiding questions and one example for each step. Figure 3 shows the impact paths that can derived from the framework including the immersive features and corresponding perception, the psychological barrier as well as individual characteristics and interventions settings as moderators. Figure 4 shows one concrete impact paths that can be derived for the example.

### 4.2.1 Step 1: Defining the Pre-Conditions

The need for behavior change can occur in various domains such as environmental behavior or health behavior and in various stages of the change process such as initiation or maintenance of new behavior. Following Wienrich & Gramlich (2020), the behavior domain (1, "Target Behavior") and the momentum within the change process (2, 3, "Intervention Settings") should be determined at the beginning of the intervention planning process. Further intervention settings should be specified, including the selection of technical setup or the need for mobility, for example. Then, the addressed target group should be defined (4, "Individual Characteristics"). Individual characteristics could be linked to the technology, such as the vulnerability to simulator sickness or the target behavior, including previous behavior change attempts (see Figure 3).

Since fully defining all pre-conditions are out of the scope of the current contribution, please see Wienrich & Gramlich (2020) or Michie et al. (2014) for detailed information about the appropriate selection of pre-conditions to design and evaluate immersive interventions. However, the following guiding questions could support the definition of pre-conditions:

1. Target Behavior: What kind of behavior does the intervention address?
2. Intervention Setting: At what moment in the behavior change process is the intervention applied?
3. Intervention Setting: In which context, how often, and when are the target persons confronted with the intervention?
4. Individual Characteristics: What are the target persons' backgrounds and characteristics, and which persons are excluded?

**Example:** The users of BehaveFIT, in our example, want to support people to engage in environmental behavior. Recently, the trend sport "plogging" (Swedish for plocka + jogging) has gained popularity. People clean their environment by collecting garbage as they run. Thus, they decided to design an immersive plogging environment to further engage people in environmentally friendly behavior. In our example, the target behavior belongs to the domain of pro-environmental behavior in general and specifically to the domain of active cleaning behavior of the local environment (refer to question 1). Target persons are already aware of environmental problems but are not actively engaged in pro-environmental behavior besides recycling (refer to question 2). The intervention will be presented in a local environment festival and people will be confronted with the intervention only once (refer to question 3). The intervention includes active physical movements as walking in place and bending over to simulate plogging, and people with physical disabilities are excluded from the intervention (refer to question 4).

### 4.2.2 Step 2: Defining the Psychological Barriers and Corresponding Immersive Features

Depending on the pre-conditions, the next step focuses on the selection of psychological barriers that target persons have to overcome. The six categories introduced in 5.1.1 guide the selection. Users should thoroughly analyze which barrier seems the most dominant and select one for a valuable intervention design (Byrne, 2020). Section 5.1.2 reveals which features and corresponding perceptions of immersive technologies can help to overcome the barriers. In most cases, various features and their corresponding perceptions can be used to address one barrier. However, for a high-quality intervention design (Byrne, 2020), developers should choose one promising immersive potential to predict concrete impact paths. To define the psychological barriers and their corresponding immersive features properly, one can use the following guiding questions (see also Figure 2):

1.  Analyze: Why do target persons show the intention-behavior-gap?
2.  Select: Which barrier category is most likely to impede the change process, and which concrete psychological barrier within the category needs to be overcome?
3.  Select: What kind of immersive feature and corresponding perception are linked to the selected barrier, and which concrete characteristic of the feature is supposed to induce the behavior change?

**Example:** In our example, the users of BehaveFIT analyzed that people who are aware but do not start engaging actively often perceive a lack of control: "What can I do?" (refer to question 1). Hence, they decided to address the barrier distance from control. Particularly, the users want to enhance the self-efficacy and the perception that their engagement matters (refer to question 2). Following the results of the successful manipulation of the self-representation, the corresponding sense of embodiment, and the Proteus effect, the users decide to design an intervention that modifies the self-representation with two different avatar-conditions. They decide to design a setting inducing a high sense of embodiment, where the users embody an athletic avatar from a first-person perspective and in which they can control the avatar's movements to be engaged in the virtual plogging experience (see Figure 4). Additionally, they design a differing condition without embodiment, in which participants observe another person engaging in the virtual plogging activity (refer to question 3).

### 4.2.3 Step 3: Defining and Evaluating the Impact Path

To control the intervention's expected impact on behavior change, users should define the impact path of their intervention. The immersive intervention should impact the corresponding perceptions (1, "Corresponding Perceptions") that in turn take effect on the psychological barrier. In the case of a successful intervention, the impact path would predict a successful behavior change or at least a facilitation of some aspects of the new behavior. In contrast, negative results show which expected impact does not occur, leading to a targeted intervention adaptation. An overview of the influence path from immersive features to behavior is given in Figure 3. It includes the immersive features, the corresponding perceptions as mediators, the psychological barriers, and the target behavior/behavior change. Additionally, it illustrates how individual characteristics can moderate the processing of immersive content and, together with the intervention settings, how they can moderate the relation between the barriers and behavior/behavior change. To define concrete impact paths, one can use the following guiding questions:

1.  Select: How should immersive features affect the corresponding perceptions?
2.  Select: How should the corresponding perceptions affect the psychological barrier?
3.  Select: How should the barrier affect behavior/behavior change?

4.    Check: Which other variables can confound the expected relations?

**Example:** In our example, the selected immersive feature is self-representation and, more specifically, a sense of control over an avatar. Thus, the users of BehaveFIT decide to modify the feature of self-representation (two avatar-conditions) that should enhance the sense of embodiment (see Figure 4), particularly the sense of control in VR (perceived agency) (refer to question 1). Furthermore, they assume a relationship between the perceived agency and the psychological feeling of self-efficacy of environmental cleaning behavior (refer to question 2). Finally, they expect that the increase of self-efficacy impacts participants' real behavior after the immersive intervention and increases active engagement in a pro-environmental activity (refer to question 3). Other confounding variables might be personal traits such as openness to new experiences (refer to question 4).

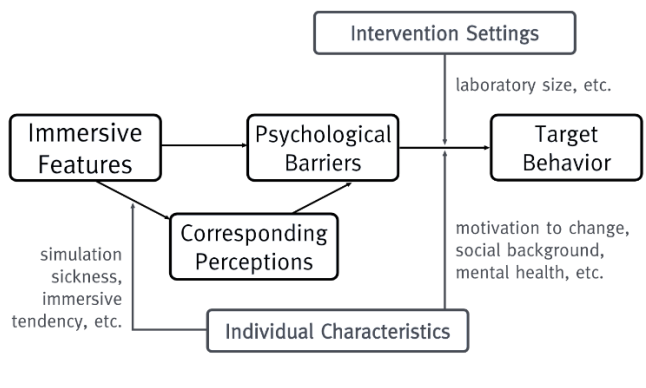
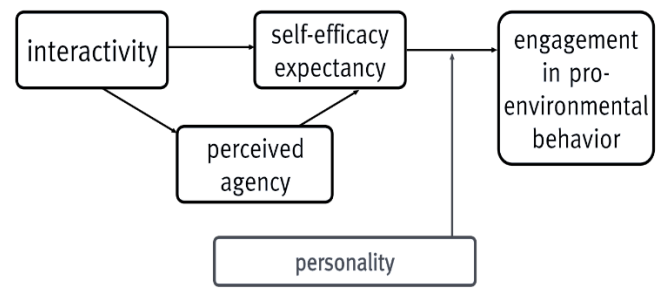

**Figure 3: Overview of the BehaveFIT impact paths showing the expected impact of immersive technologies, moderators, and situational factors on behavior change.**

**Figure 4: One possible impact paths illustrating the expectations of the example.**

## 5    Discussion

### 5.1    Review Of Present Work

The intention-behavior-gap describes the failure of the translation from intending to change a behavior to successfully changing it (Ajzen, 2011; Carrington et al., 2010; Gollwitzer & Bargh, 1996; Kollmuss & Agyeman, 2002; Schwarzer, 2014; Sheeran, 2002; Whitmarsh, 2009). The related work revealed plenty of research on psychological barriers describing why the gap occurs and behavioral models explaining how behavioral interventions can overcome the gap. Furthermore, features of immersive interventions explain why they have a lot of potential to support behavior change processes and empirical findings indicating how they impact human behavior. However, the literature review also revealed four desiderata corroborating the need for an intelligible integration of psychological barriers, models, and immersive features. The current lack of transfer between the disciplines results in a lack of theoretical links between the features and possibilities of immersive technologies and psychological behavior determinants, which further results in a lack of theoretical-based predictions for the design and evaluation of immersive interventions. The framework BehaveFIT addresses the desiderata by (see also Figure 2):

- presenting an intelligible categorization of psychological determinants and barriers. The psychological barriers are described in six distances impeding the detachment of old behavior or the adoption of new behavior (desideratum 1).
- suggesting a mapping of immersive features and corresponding perceptions on to the six distances (barrier categories). The mapping shows why and where immersive technologies can help overcome the barriers (desideratum 3).
- addressing the summation of psychological models on how behavior change techniques operate and research how immersive technologies affect behavior change (desiderata 2).
- proposing a generic prediction path that enables a structured, theory-based approach to develop and evaluate immersive interventions (desideratum 4). There are three steps, which explain how BehaveFIT can be used, including guiding questions and one example for each step.

## 5.2  Benefits and Contribution of the BehaveFIT

Byrne (2020) identifies the most critical point in his review about gaps and priorities in advancing methods for health behavior change research is the consideration of behavioral change theories and models. Further, and in accordance with Wobbrock & Kientz (2016), a theory-driven approach has the effect of finding out not only if an intervention works but also why and how it works. This leads to evaluation results that are more generally applicable to further developments and new interventions in other areas. In addition, it supports the development of more targeted interventions by drawing on mechanisms already known about from psychological research. The present paper seized on this challenge by suggesting an interdisciplinary cross transfer to provide why and how immersive interventions support behavior change processes. BehaveFIT goes beyond existing suggestions as to why and how immersive intervention technologies might support behavioral change. The proposed relations between the psychological barriers and the immersive features serve as a starting point for predicting concrete paths of impact on immersive interventions. The summation, categorization, and intelligible mapping offer systematic guidance for design effectual immersive interventions. Concerning human-centered design, the pre-conditions of the interventions can be analyzed (step 1), including the context, target group, and confounding variables. Then, the psychological barriers and the corresponding immersive features can be defined in a structured and theory-driven way (step 2).

Concerning evaluation, Klasnja et al. (2011) suggested that assessing whether a technology accomplished behavioral change often leads to limited results. Due to the complexity of behavior change, evaluations "[...] often do not reveal why the technology that is being evaluated worked or did not work" (Klasnja et al., 2011, p. 3064). The authors stated further: "This knowledge, however, is of central importance to HCI researchers as they attempt to develop novel, yet effective systems" (Klasnja et al., 2011, p. 3064). BehaveFIT can contribute here in various ways. The predictions resulting from the concrete impact paths reduce the corresponding evaluation's complexity and thus are able to reveal why the technology worked or did not work (step 3). In line with Klasnja et al. (2011), we believe that knowing why is crucial to HCI researchers as they attempt to develop effective systems. Furthermore, since the framework includes established factors (psychological and technological), we recommend researchers following Wienrich & Gramlich (2020) to consider measurements that account for corresponding measurements assessing the impact or rather the manifestations of these effects (behavioral effects: e.g., stage of change questionnaire (Marcus et al., 1992) or index of habit strength (Verplanken & Orbell, 2003); immersive features and perceptions: e.g., virtual embodiment questionnaire (Roth & Latoschik, 2020),     group presence questionnaire (Schubert et al., 2001) or networked minds measure of social presence (Biocca & Harms, 2003).

In sum, BehaveFIT provides guidance for immersive intervention design and evaluation to show that immersive interventions support behavior change processes and explain and predict why and how immersive interventions can bridge the intention-behavior-gap.

## 5.3    Limitations and Future Work

This framework is the first attempt at a structured integration of potentials arising from immersive technologies and psychological barriers determining behavioral change processes. The current framework can give a broad overview of the connection between these two fields. Consequently, the framework comes with certain restrictions concerning levels of detail of the framework's components, the impact paths, and application areas.

### 5.3.1 Further Components

Due to summation and comprehensibility, we accepted some restrictions regarding the complexity of mental processes and determinants during the behavioral change. Only a selection of mental inhibitors and facilitators have been included, although all of them can be divided into various sub-facets or sub-sub-facets (for a review, see Cane et al., 2012). The same limitations pertain to the factors of the immersive features and corresponding perceptions. Here, too, we would like to point out that the framework is intended as a selection to provide intelligible summation and guidance, and it does not claim completeness.

Besides including further barriers and features, future work might focus on identifying and weighing the most critical determinants influencing behavior change progress. Weighting the barriers and features might help to get further insights into the relevance or irrelevance of the determinants, which in turn might facilitate the guidance for the design and evaluation of immersive interventions.

### 5.3.2 Further Paths of Impact

Even though users of the current framework can extract specific paths of impact by mapping immersive features and psychological barriers, we do not define one-to-one relations between the features and barriers. As shown in the literature review, the current state of research does not yet allow such an exact assignment. Thus, the framework instead serves as a reference guide in HCI research for the design and evaluation of immersive intervention technologies. However, by investigating specific impact paths and interrelations between the barriers and features in future work, it will gradually be possible to include them in the framework structure.

### 5.3.3 Further Fields of Application

Since the majority of research activity has been done on virtual reality applications, the included immersive features and corresponding perceptions are based on VR. Hence, the features might not be necessarily transferable to other (immersive) technologies such as augmented reality (AR) or 360°-videos. Future work could examine the immersive dimensions in the context of non-VR-related technologies.

In sum, further development of BehaveFIT 1.0 might include both other therapeutic application areas and other immersive or partly immersive technologies. Further research might lead to a respective enlargement of the current framework and the addressed areas of application.

## 5.4    Conclusion

When choosing to change a specific behavior, people often cannot attain their goals, leading up to a gap between behavioral intention and actual behavior. Within HCI research, the development and evaluation of immersive technologies that target supporting the behavior change process and bridging this gap have gained importance. However, a significant part of these developments still lacks the theoretical background or remain limited to the initiation of a behavior or to short-term effects. This paper presented a first attempt to combine two research areas – the field of psychology examining behavior change and the field of HCI investigating immersive technologies and the corresponding human sensations and perceptions – in a comprehensive framework. BehaveFIT wants to guide the design and evaluation process of immersive intervention tools by assisting in the contextual analysis, including personas.

Additionally, it may provide guidance in identifying, inhibiting or facilitating mediators between immersive technologies and behavioral measures. A review of HCI research revealed that only a few determinants or paths of impact had been attended so far. Hence, we believe that the present work is an essential contribution to the encouragement of researchers from both fields to use BehaveFIT to bridge the gaps.

## 6 References

Ahn, S. J., Bailenson, J. N., & Park, D. (2014). Short- and long-term effects of embodied experiences in immersive virtual environments on environmental locus of control and behavior. *Computers in Human Behavior 39*, pp. 235–245.

Ahn, S. J., Bostick, J., Ogle, E., Nowak, K., McGillicuddy, K., & Bailenson, J. (2016). Experiencing Nature: Embodying Animals in Immersive Virtual Environments Increases Inclusion of Nature in Self and Involvement With Nature. *Journal of Computer-Mediated Communication 21(6)*, pp. 399–419.

Ahn, S. J., Fox, J., Dale, K. R., & Avant, J. A. (2015). Framing Virtual Experiences: Effects on Environmental Efficacy and Behavior over Time. *Communication Research 42(6)*, pp. 839–863.

Ahn, S. J., Hahm, J. M., & Johnsen, K. (2019). Feeling the weight of calories: using haptic feedback as virtual exemplars to promote risk perception among young females on unhealthy snack choices. *Media Psychology 22(4)*, pp. 626–652.

Ajzen, I. (1980). Understanding attitudes and predicting social behavior. *Englewood Cliffs.*

Ajzen, I. (2011). The theory of planned behaviour: Reactions and reflections.

Bandura, A. (1977). Self-efficacy: toward a unifying theory of behavioral change. *Psychological review 84(2)*, pp. 191.

Baumeister, R. F., & Leary, M. R. (1997). Writing narrative literature reviews. *Review of general psychology 1(3)*, pp. 311–320.

Becker, M., & Maiman, L. (1975). Sociobehavioral determinants of compliance with health and medical care recommendations. *Medical care*, pp. 10–24.

Biocca, F., & Harms, C. (2002). Defining and measuring social presence: Contribution to the networked minds theory and measure. *Proceedings of PRESENCE, 2002*, pp. 1–36.

Biocca, F., & Harms, C. (2003). Guide to the Networked Minds Social Presence Inventory v. 1.2: Measures of co-presence, social presence, subjective symmetry, and intersubjective symmetry. *Michigan State University, East Lansing*.

Blake, J. (1999). Overcoming the ,value-action gap 'in environmental policy: Tensions between national policy and local experience. *Local environment 4(3)*, pp. 257–278.

Bolger, N., & Zee, K. S. (2019). Heterogeneity in temporal processes: Implications for theories in health psychology. *Applied Psychology: Health and Well-Being 11(2)*, pp. 198–201.

Bridle, C., Riemsma, R. P., Pattenden, J., Sowden, A. J., Mather, L., Watt, I. S., & Walker, A. (2005). Systematic review of the effectiveness of health behavior interventions based on the transtheoretical model. *Psychology & Health 20(3)*, pp. 283–301.

Byrne, M. (2020). Gaps and priorities in advancing methods for health behaviour change research. *Health Psychology Review 14(1)*, pp. 165–175.

Cane, J., O'Connor, D., & Michie, S. (2012). Validation of the theoretical domains framework for use in behaviour change and implementation research. *Implementation science 7(1)*, pp. 37.

Carrington, M., Neville, B., & Whitwell, G. (2010). Why ethical consumers don't walk their talk: Towards a framework for understanding the gap between the ethical purchase intentions and actual buying behaviour of ethically minded consumers. *Journal of business ethics 97(1)*, pp. 139–158.

Davis, R., Campbell, R., Hildon, Z., Hobbs, L., & Michie, S. (August 2014). Theories of behaviour and behaviour change across the social and behavioural sciences: a scoping review. *Health Psychology Review 9(3)*.

De Kort, Y., IJsselsteijn, W., & Poels, K. (2007). Digital games as social presence technology: Development of the Social Presence in Gaming Questionnaire (SPGQ). *Proceedings of PRESENCE 195203*.

Fox, J., & Bailenson, J. (2009). Virtual self-modeling: The effects of vicarious reinforcement and identification on exercise behaviors. *Media Psychology 12(1)*, pp. 1–25.

Gifford, R., Lacroix, K., & Chen, A. (2018). Understanding responses to climate change: Psychological barriers to mitigation and a new theory of behavioral choice. *In Psychology and Climate Change, Elsevier, Academic Press,* pp. 161–183.

Gollwitzer, P., & Bargh, J. (1996). The psychology of action: Linking cognition and motivation to behavior. *Guilford Press*.

Hassenzahl, M., Diefenbach, S., & Göritz, A. (2010). Needs, affect, and interactive products–Facets of user experience. *Interacting with computers 22(5)*, pp. 353–362.

Herrera, F., Bailenson, J., Weisz, E., Ogle, E., & Zaki, J. (2018). Building long-term empathy: A large-scale comparison of traditional and virtual reality perspective-taking. *PloS one 13(10)*, pp. e0204494.

Holtzblatt, K., & Beyer, H. (2017). Principles of Contextual Inquiry. *Contextual Design*, pp. 43-80.

Hsu, W.-C., Tseng, C.-M., & Kang, S.-C. (2018). Using Exaggerated Feedback in a Virtual Reality Environment to Enhance Behavior Intention of Water-Conservation. *Educational Technology & Society 21*, pp. 187–203.

Kilteni, K., Groten, R., & Slater, M. (2012). The sense of embodiment in virtual reality. *Presence: Teleoperators and Virtual Environments 21(4)*, pp. 373–387.

Kim, S., Prestopnik, N., & Biocca, F. (2014). Body in the interactive game: How interface embodiment affects physical activity and health behavior change. *Computers in Human Behavior 36 (2014)*, pp. 376–384.

Klasnja, P., Consolvo, S., & Pratt, W. (2011). How to evaluate technologies for health behavior change in HCI research. *In Proceedings of the SIGCHI conference on human factors in computing systems. ACM*, pp. 3063–3072.

Kollmuss, A., & Agyeman, J. (2002). Mind the gap: why do people act environmentally and what are the barriers to pro-environmental behavior? *Environmental education research 8(3)*, pp. 239–260.

Kuniavsky, M. (2003). *Observing the user experience: a practitioner's guide to user research.* Elsevier.

Kuo, P.-Y. (April 2015). Engage People in Pro-Environmental Behaviors through Online Prosocial Interaction and Pro-Health Intervention. *In Proceedings of the 33rd Annual ACM Conference Extended Abstracts on Human Factors in Computing Systems - CHI EA '15*, pp. 207-210.

Maister, L., Slater, M., Sanchez-Vives, M. V., & Tsakiris, M. (2015). Changing bodies changes minds: owning another body affects social cognition. *Trends in cognitive sciences*, 19(1), 6-12.

Marcus, B., Selby, V., Niaura, R., & Rossi, J. (1992). Self-efficacy and the stages of exercise behavior change. *Research quarterly for exercise and sport 63(1)*, pp. 60–66.

Michie, S., Atkins, L., & West, R. (2014). The behaviour change wheel. A guide to designing interventions. *1st ed. Great Britain: Silverback Publishing*, pp. 1003–1010.

Michie, S., Richardson, M., Johnston, M., Abraham, C., Francis, J., Hardeman, W., . . . Wood, C. (2013). The behavior change technique taxonomy (v1) of 93 hierarchically clustered techniques: building an international consensus for the reporting of behavior change interventions. *Annals of behavioral medicine 46(1)*, pp. 81–95.

Moher, D., Shamseer, L., Clarke, M., Ghersi, D., Liberati, A., Petticrew, M., . . . Stewart, L. (2015). Preferred reporting items for systematic review and meta-analysis protocols (PRISMA-P) 2015 statement. *Systematic reviews 4(1)*, pp. 1.

Morina, N., Ijntema, H., Meyerbröker, K., & Emmelkamp, P. (Katharina Meyerbröker, and Paul MG Emmelkamp 2015). Can virtual reality exposure therapy gains be generalized to real-life? A meta-analysis of studies applying behavioral assessments. *Behaviour research and therapy 74*, pp. 18–24.

Oh, C., Bailenson, J., & Welch, G. (2018). A systematic review of social presence: Definition, antecedents, and implications. *Frontiers in Robotics and AI, (5)*, pp. 114.

Powers, M., & Emmelkamp, P. (2008). Virtual reality exposure therapy for anxiety disorders: A meta-analysis. *Journal of anxiety disorders, 22(3)*, pp. 561-569.

Prochaska, J., & Velicer, W. (1997). The transtheoretical model of health behavior change. *American journal of health promotion, 12(1)*, pp. 38-48.

Ratan, R., Beyea, D., Li, B., & Graciano, L. (2020). Avatar characteristics induce users' behavioral conformity with small-to-medium effect sizes: a meta-analysis of the proteus effect. *Media Psychology, 23(5)*, pp. 651-675.

Riva, G., Wiederhold, B., & Mantovani, F. (2019). Neuroscience of virtual reality: from virtual exposure to embodied medicine. *Cyberpsychology, Behavior, and Social Networking, 22(1)*, pp. 82-96.

Roth, D., & Latoschik, M. (2020). Construction of the Virtual Embodiment Questionnaire (VEQ). *IEEE Transactions on Visualization and Computer Graphics*.

Rothman, A. (2019). It's About Time: Answering the Call for Greater Precision in Research and Practice. *Applied Psychology: Health and Well-Being, 11(2)*, pp. 191-197.

Scholz, U. (2019). It's time to think about time in health psychology. *Applied Psychology: Health and Well-Being, 11(2)*, pp. 173-186.

Schubert, T., Friedmann, F., & Regenbrecht, H. (2001). The experience of presence: Factor analytic insights. *Presence: Teleoperators & Virtual Environments, 10(3)*, pp. 266-281.

Schwarzer, R. (2014). *Self-efficacy: Thought control of action.* Taylor & Francis.

Sheeran, P. (2002). Intention-behavior relations: a conceptual and empirical review. *European review of social psychology, 12(1)*, pp. 1-36.

Sherman, W., & Craig, A. (2018). *Understanding virtual reality: Interface, application, and design.* Morgan Kaufmann.

Skarbez, R., Brooks, J., & Whitton, M. (2017). A survey of presence and related concepts. *ACM Computing Surveys (CSUR), 50(6)*, pp. 1-39.

Slater, M. (1999). Measuring presence: A response to the Witmer and Singer presence questionnaire. *Presence 8, 5 (1999)*, pp. 560–565.

Slater, M. (2003). A note on presence terminology. *Presence connect, 3(3)*, pp. 1-5.

Soliman, M., Peetz, J., & Davydenko, M. (2017). The impact of immersive technology on nature relatedness and pro-environmental behavior. *Journal of Media Psychology*.

Stawarz, K., & Cox, A. (2015). Designing for health behavior change: HCI research alone is not enough. *In CHI'15 workshop: Crossing HCI and Health: Advancing Health and Wellness Technology Research in Home and Community Settings*.

Stawarz, K., Cox, A., & Blandford, A. (2015). Beyond self-tracking and reminders: designing smartphone apps that support habit formation. *In Proceedings of the 33rd annual ACM conference on human factors in computing systems*, pp. 2653-2662.

Verplanken, B., & Orbell, S. (2003). Reflections on past behavior: a self-report index of habit strength. *Journal of applied social psychology, 33(6)*, pp. 1313-1330.

Whitmarsh, L. (2009). Behavioural responses to climate change: Asymmetry of intentions and impacts. *Journal of environmental psychology 29(1)*, pp. 13–23.

Wienrich, C., & Gramlich, J. (2020). appRaiseVR–An Evaluation Framework for Immersive Experiences. *i-com, 19(2)*, pp. 103-121.

Wienrich, C., Schindler, K., Döllinger, N., Kock, S., & Traupe, O. (2018). Social presence and cooperation in large-scale multi-user virtual reality-the relevance of social interdependence for location-based environments. *In 2018 IEEE Conference on Virtual Reality and 3D User Interfaces (VR). IEEE*, pp. 207–214.

Wilson, T., Hennessy, E., Falzon, L., Boyd, R., Kronish, I., & Birk, J. (2020). Effectiveness of interventions targeting self-regulation to improve adherence to chronic disease medications: A meta-review of meta-analyses. *Health Psychology Review 14(1)*, pp. 66–85.

Wobbrock, J., & Kientz, J. (2016). Research contributions in human-computer interaction. *interactions 23(3)*, pp. 38–44.

Yee, N., & Bailenson, J. (2007). The Proteus effect: The effect of transformed self-representation on behavior. *Human communication research 33(3)*, pp. 271–290.